\documentclass[12pt,preprint]{emulateapj}

\shortauthors{Hsieh \& Yee}

\begin{document}

\title{Estimating Luminosities and Stellar Masses of Galaxies
Photometrically without Determining Redshifts}
\author{B. C. Hsieh\altaffilmark{1},
H. K. C. Yee\altaffilmark{2}
}

\altaffiltext{1}{Institute of Astronomy \& Astrophysics, Academia Sinica,
P.O. Box 23-141, Taipei 106, Taiwan, R.O.C. Email: bchsieh@asiaa.sinica.edu.tw}

\altaffiltext{2}{Department of Astronomy \& Astrophysics, University of
Toronto, Toronto, Ontario, M5S 3H4, Canada. Email: hyee@astro.utoronto.ca}

\begin{abstract}
Large direct-imaging surveys usually use a template-fitting technique to estimate
photometric redshifts for galaxies, which are then applied to derive important galaxy
properties such as luminosities and stellar masses.
These estimates can be noisy and suffer from systematic biases because of
the possible mis-selection of templates and the propagation of the
photometric redshift uncertainty. We introduce an algorithm, the Direct
Empirical Photometric method (DEmP), which can be used to directly estimate
these quantities using training sets, bypassing photometric redshift determination.
DEmP also applies two techniques to minimize the effects arising from the
non-uniform distribution of training-set galaxy redshifts from a flux-limited
sample. First, for each input galaxy, fitting is performed
using a subset of the training-set galaxies with photometry and colors closest to
those of the input galaxy. Second, the training set is artificially resampled
to produce a flat distribution in redshift, or other properties, e.g.,
luminosity. To test the performance of DEmP, we use
a 4-filter-band mock catalog to examine its ability to recover
redshift, luminosity, stellar mass, and
luminosity and stellar-mass functions. We also compare the results to those
from two publicly available template-fitting methods, 
finding that the DEmP algorithm outperforms both. We find
resampling the training set to have a uniform redshift distribution
produces the best results not only in photometric redshift, but
also in estimating luminosity and stellar mass.
The DEmP method is especially powerful in estimating quantities such as
near-IR luminosities and stellar mass using only data from a small
 number of optical bands.
\end{abstract}


\section{Introduction}\label{introduction}
Luminosity and stellar mass are important physical properties of a galaxy.
They are also used for selecting galaxy samples for many extragalactic studies
(e.g., luminosity-limited or stellar-mass-limited samples).
How to estimate these quantities accurately for very large data sets,
such as wide-field imaging surveys, is therefore crucial.
Multi-broad-band imaging is an efficient way to obtain
large and statistically significant samples of galaxies.
Quantities such as luminosity, colors, and stellar mass can be
derived from these data, usually using a spectral-energy-distribution
(SED) fitting method.
For a large surveys, where a large sample spectroscopic redshift sample is
difficult to obtain, photometric redshifts derived from the same
photometry are often used in deriving these quantities.
In this case, the performance of the photometric redshift method, both in terms
of accuracy and bias, is crucial to the estimate of quantities such as
luminosity and stellar mass, as uncertainties in the photometric redshifts
are propagated and magnified when estimating these quantities.

Methods for deriving photometric redshifts can be generally
grouped into two categories:
template fitting \citep[e.g.,][]{arnouts1999,bmp2000,benitez2000,brammer2008}
and empirical fitting \citep[e.g.,][]{connolly1995,firth2003,hsieh2005}.
The template-fitting method fits the galaxy SED
with templates constructed from either real spectrophotometric data
\citep[e.g.,][]{cww1980}
or stellar population synthesis models \citep[e.g.,][]{bc2003},
or a combination of both.
The empirical-fitting method assumes that the redshift of a galaxy
is a function of its photometry (e.g., redshift $= f_z(g,r,i',z')$).
It utilizes a spectroscopic training set
to derive this function and then applies this function to
the photometry for all the objects to compute their redshifts.
In a recent paper, \citet{hildebrandt2010} summarized 
the performance of a number of existing photometric redshift algorithms.

The performance of the template photometric redshift methods
depends heavily on the chosen templates.
If the templates are not representative of the SEDs of real objects,
the results could be biased or incorrect.
This issue can be minimized if more than a dozen broad-
and intermedian-bands of photometry are available 
\citep[e.g., COSMOS 30-band photometry,][]{ilbert2009}
which can provide adequate constraints to determine a proper template
for each galaxy and avoid degeneracies between different templates.
However, it is difficult to collect photometry for more than 10 bands
in a large survey field (e.g., larger than 100 square degrees);
photometry with only a few bands (e.g., 4 or 5) 
generally provides poorer constraints in SED fitting
which increases the chance of picking up a wrong template for a given galaxy.

Unlike the template methods, the empirical methods estimate redshifts 
based on the empirical relation between redshifts and photometry.
One therefore does not need to apply any template,
as long as the training set is complete over the selected redshift range.
For redshift less than 1.5,
publicly available databases released by 
several deep spectroscopic redshift surveys 
\citep[e.g.,][]{lefevre2005,lilly2007,davis2007}
provide excellent samples to generate high-quality training sets
for the empirical photometric redshift method;
the training sets constructed using these survey data 
do not suffer from significant completeness issues for $z < 1.5$.
Because there is no templates involved in an empirical method,
it can avoid introducing additional noises, biases,
and uncertainties due to mis-selected templates,
and thus could result in better photometric redshift quality.
This is the most important advantage of empirical methods
over template methods.
However, note that this method requires addtional data 
to create the training set.

A similar situation occurs when estimating luminosities and stellar masses
of galaxies using a template-fitting method.
This method would suffer all the issues
found in template photometric redshift methods.
Furthermore, the noise and bias of the photometric redshift
used for deriving the luminosity and the stellar mass
are propagated to the final result and
could produce an even more deleterious effect.
This is especially serious in data with only a small number of filter bands.
If luminosities and stellar masses can be estimated  
bypassing the redshift determination, these issues can be minimized.
\citet{firth2003} show that the morphology of galaxy can be estimated
from photometry empirically, without deriving their redshifts.
\citet{hsieh2008} derive the $k$-correction of galaxy from photometry directly
without involving redshifts.
\citet{budavari2009} introduces a unified framework
in which the photometric redshift method is simply a mapping of
the photometry/color space to the redshift space.
Furthermore, other physical properties can also be estimated directly by
mapping the photometry/color space to the spaces of these properties.
In this paper, we develop a simple, but powerful and robust, empirical
method which, besides being applicable to deriving photometric redshift,
can be used to estimate luminosity and stellar mass directly using photometry
from a relatively small number of filter bands, bypassing the determination
of photometric redshift completely.
We examine its performance using a mock photometric catalog of 4 optical bands 
and compare the results to those obtained using 
the more traditional template methods.

This paper is structured as follows. 
In \S\ref{method} we describe the empirical method
that we use to estimate redshifts, luminosities, and stellar masses.
In \S\ref{mock} we provide a description of the mock catalog 
that we use to examine the performance of our empirical method.
Section \ref{trainingset} presents 
the construction of the training sets and the validation sets
that are used in our experiments.
We then compare the qualities of photometric redshifts, luminosities,
and stellar masses estimated using our empirical method and
several conventional template-fitting methods in \S\ref{results},
and discuss the results in \S\ref{discussion}. 
In \S\ref{summary} we summarize our results. 
The cosmological parameters used in this study are 
$\Omega_\Lambda = 0.7$, $\Omega_M = 0.3$, $H_0 = 70$ km s$^{-1}$Mpc$^{-1}$, 
and $w = -1$. 

\section{The Direct Empirical Photometric Method}\label{method}
The idea of using broad-band photometric data
to estimate galaxy luminosities and stellar masses directly 
using a training set is simple.
The empirical photometric redshift method
assumes that the redshift of a galaxy is a function of its photometry.
Similarly, if one simply assumes that the luminosity 
and stellar mass of a galaxy
are functions of its photometry (i.e., luminosity $= f_L(m_1, m_2, ..., m_N)$,
and stellar mass $= f_M(m_1, m_2, ..., m_N)$, 
where $m_i$ is the magnitude for the $i$th filter),
then the luminosities and stellar masses of galaxies can be derived
directly from their photometry,
without involving any redshifts and templates.
The training sets for the empirical luminosity and stellar-mass fittings
can be constructed using publicly available deep spectroscopic 
and multi-wavelength photometric datasets
\citep[e.g.,][]{lefevre2005,lilly2007,davis2007}.
When the luminosities and stellar masses of the training set galaxies 
are derived using SED fitting with photometry from 
more than a dozen broad- and intermediate-bands along with spectroscopic redshifts,
the mis-selection of templates can be mostly avoided.
The training sets for luminosity and stellar mass
constructed as described are thus likely reliable.
Alternatively, for the stellar-mass training set,
other probes can also be utilized to derive 
the stellar masses in the training set
\citep[e.g., stellar mass versus $K$-band luminosity relation,][]{be2000}
if they are deemed more suitable than the SED fitting method.
Note that this empirical-fitting formalism can be applied
to any intrinsic properties, such as $k$-corrections \citep[e.g.,][]{hsieh2008}
and rest-band colors ($m_i$--$m_j$), that are tied to the observed photometry.

The conventional empirical-fitting methods, however, 
have two major issues that could affect the result significantly.
One is the choice of the proper form of the fitting functions,
e.g., $f_z$, $f_L$, and $f_M$, etc.
The other is that the best-fitted coefficients for the empirical functions
can be biased by objects with higher population density
(e.g., the mid-redshift population for $f_z$, 
the low luminosity population for $f_L$,
and the low stellar-mass population for $f_M$).
This effect is also discussed in \citet{budavari2009}.
Our new empirical-fitting method is designed
to minimize the effects of these issues.
In the following subsections,
we describe how these two issues are dealt with
in our empirical photometric redshift method.
While we discuss these techniques in terms of deriving photometric
redshifts, they can be generalized to the direct empirical methods
for deriving other galaxy properties
such as luminosity and stellar mass, and we discuss their
implementations and tests in subsequent sections.

\subsection{Regional Polynomial Fitting}\label{poly}
Finding a proper form of the empirical function
is critical for the empirical methods.
\citet{connolly1995} use a polynomial form and
find that the higher order the polynomial form,
the smaller the fitting $\chi^2$.
However, the $\chi^2$ is only reduced slightly
when they move from a third- to a fourth-order fit.
They suggest that this is because the curvature 
in the photometry-versus-redshift relation
is not in all the dimensions of the magnitude-color-redshift space.
However, an alternate explanation could be that
a polynomial form may not properly describe the relation exactly.
Alternatively, the code {\it Artificial-Neural-Network-z} 
\citep[ANNz,][]{firth2003}
takes a more aggressive approach,
in which an artificial neural network is used for deriving
the empirical relation between redshift and photometry.
The more the layers (or nodes) there is in the network 
(i.e., more complicated network),
the better the photometric redshift performance.
However, once the complexity of the network reaches a certain degree,
the improvement becomes negligible,
which is similar to the afore-mentioned issue of the polynomial fitting method.
These findings suggest that
even using a very complicated form for the empirical function
does not help significantly in improving 
the quality of the photometric redshift.

Since even complicated functions and algorithms have limitations
in delivering better photometric redshifts, 
we can opt for a simple approach of fitting the function piece-wise.
In \citet{hsieh2005}, we divide up the data points
into several partitions in color-magnitude space
using the $kd$-tree algorithm \citep{bentley1979},
and derive the empirical relation in each partition
using a second-order polynomial function.
A complicated curve can be mimicked using multiple line segments;
similarly, deriving the empirical relation
for each color-magnitude partition individually
using a simple low-order polynomial function 
may achieve a better performance when compared to a brute-force single fit 
with a very complicated function for all the data.
For the partitional photometric redshift method,
the more partitions there are, the better the performance.
Ideally, it is better to divide up the data points
on all the axes of colors and magnitudes;
however, the number of galaxies in a typical training set
is likely not adequate for such an operation.
For example, a training set of 10,000 galaxies
with four broad-band photometry can inhabit
a 10-dimensional color-magnitude space (6 unique colors and 4 magnitudes).
If one wants to divide up the data points 
on all the color and magnitude axes once at least,
there will be more than one thousand partitions.
This would mean that there are on average 
less than 10 galaxies in each partition.
Thus, one would be in danger of overfitting
because of having too few objects per partition.
The typical number of galaxies of a high-quality training set
that can be generated nowadays is about 10,000 to 20,000,
similar to that used in our training set.
To avoid the overfitting issue, another approach has to be taken.

Instead of dividing the training-set galaxy sample into many fixed partitions,
an alternate method is to select a sufficiently large subset of the galaxies,
e.g., 50, whose magnitudes and colors are closest to the input galaxy,
a ``local subset''.
This dynamically-selected local subset is used for 
deriving the relation between redshift and photometry
for that input galaxy, with a first-order polynomial function
\footnote {A higher-order polynomial function can be used here,
but the number of galaxies in each local subset has to be increased 
to avoid over-fitting, 
which also means an increase in computational time.
In addition, fitting with a higher-order polynomial function
does not guarantee better results
compared to those obtained using a first-order one,
and sometimes this introduces artifacts.
We therefore choose a first-order polynomial function
with a small but sufficiently large local subset 
(50 galaxies) for the fitting.}.
Similar methods are also used in \citet{csabai2007} and \citet{li2012}.
In our method, the quadratically summed ranks of color and magnitude differences
between the training set galaxies and the input galaxy
are used for constructing the local subset.
In addition, we also assign a weight to each galaxy in the local subset,
based on the inverse-cube value of the distance
between that galaxy in the subset and the input galaxy
in the multi-dimensional color-magnitude space.
An important factor in ranking the training set galaxies
is that the ranges of the distributions 
in different galaxy colors and magnitudes are very different,
which can bias the selection of galaxies 
for the local subset.
For example, the magnitude distributions of a catalog
with $g$, $r$, $i'$, $z'$ photometry
may range from 15 to 25 mag (i.e., a 10-mag dynamical range),
but the color distributions likely have dynamical ranges of $<3$ mag.
This means that a local subset with equal weights in color and magnitude distances
will tend to have galaxies with similar magnitudes
rather than those with similar colors,
which would degrade the photometric redshift performance,
since colors usually provide more redshift information than magnitudes.
Therefore, a scaling factor is needed to be applied to each color and magnitude
when calculating the distance in the multi-dimensional space
to optimize the selection of galaxies.
The initial values of the scaling factors can be roughly determined
based on the ratio of the ranges of colors and magnitudes.
For example, one can apply a scaling factor of 3 to each color
and 1 to each magnitude,
then the ranges of all the colors and magnitudes would be similar.
In addition, the range of a given color is redshift-dependent,
and different colors have different ranges at the same redshift.
Applying different scaling factors to different colors
can therefore fine-tune the photometric redshift performance
at different redshift ranges.

To derive the photometric redshift value and its uncertainty for each galaxy, 
we follow the method described in \citet{hsieh2005}. 
To account for the effects due to photometric uncertainties, 
we use Monte Carlo technique to generate 500 data sets 
based on the photometric measurements and uncertainties of the input galaxies.
For each of the Monte Carlo generated data set, 
we bootstrap the training set for each input galaxy 500 times 
to estimate the sampling effect in the training set.
This produces 250,000 photometric redshift estimates for each input galaxy.
The median value of these 250,000 estimates is assigned to be
the photometric redshift of the galaxy,
and the distribution of these 250,000 estimates represents
the probability function of the photometric redshift 
(i.e., the photometric redshift error) for that galaxy.

We apply the same Monte Carlo plus bootstrap method for deriving 
luminosities and stellar masses of the galaxies;
the probability functions of luminosity and stellar mass
are therefore provided as well.
We give our method a general name:
the Direct Empirical Photometric method, or DEmP.

\subsection{Uniformly-Weighted Training Set}\label{balance}
All data fitting procedures are subject to biases 
due to objects with higher population density in the relevant parameter space.
For example, in a redshift training set
there are usually more objects at intermediate redshift range 
than at both the high- and low-redshift ends
because of the smaller survey volume at low redshifts
and the brighter absolute magnitude limits at high redshifts
for an apparent magnitude-limited sample.
Therefore, the best-fitting coefficients are optimized 
for the intermediate redshift range;
but low-redshift objects would have overestimated photometric redshifts
and high-redshift objects would have underestimated photometric redshifts.
If one derives photometric redshifts for a training set
using the training set itself,
the photometric redshift distribution would always be narrower
than the spectroscopic redshift distribution
because of this bias effect.
Using a partitional or regional fitting method 
as describe in Section~\ref{poly} might reduce this effect
if the total number of objects in the training set are adequate
(e.g., more than 100,000 objects for $z < 1.2$, according to our experience).
In this case, 
since objects in each partition are expected to have similar redshifts,
the redshift distribution in each partition would be approximately ``flat''.
However, the best training set that can be constructed nowadays 
includes only around 20,000 objects, 
making a partition method still affected by this bias effect.  
Therefore, some other steps have to be taken to alleviate this problem.

The most intuitive solution is to give a weight inversely proportional 
to the population density function to each object in the training set.
This is similar to the uniform weighting
used in data reduction in radio astronomy.
We adopt another straightforward approach by artificially making 
the number of galaxies in each redshift bin in the training set the same.
The new training set would therefore have a flat redshift distribution.
We call this training set a ``uniformly-weighted'' training set.
For regional polynomial fitting, ideally, 
this procedure should be performed for each local subset of the training set.
However, such an algorithm would be very computing-time intensive.
In fact, based on our tests, we have found a negligible quality difference
between using several uniformly-weighted subsets
and using a single uniformly-weighted training set.
Therefore, we use a single uniformly-weighted training set in DEmP,
without compromising the quality of the result.
The details for the construction of the uniformly-weighted training set
are described in Section~\ref{trainingset}.
 
\section{RCS2 Mock Catalog}\label{mock}

To examine the performance of DEmP,
we generated a mock catalog that mimics the data of
the second Red-Sequence Cluster Survey
\citep[RCS2,][]{gilbank2011}.
RCS2 is a multi-band imaging survey covering nearly 1,000 square degrees
carried out using the square-degree imager, 
MegaCam, on CFHT.
It is designed to search for clusters of galaxies 
over the redshift range between 0.1 and 1.0.
The project uses the red-sequence of cluster early-type galaxies
to identify clusters \citep{gy2000}.
The survey comprises three-filter imaging ($g$, $r$, and $z'$),
with additional $i'$-band imaging via a data-exchange
with the Canada-France High-$z$ Quasar Survey \citep{willott2005}.
The 5$\sigma$ limiting magnitudes in AB are 24.4, 24.3, 23.7, and 22.8, 
for $g$, $r$, $i'$, and $z'$, respectively.
The calibrated RCS2 photometry has an absolute accuracy of better than 0.03 mag
on any color and $\sim0.05$ mag in the $r$-band magnitude,
verified with respected to the Sloan Digital Sky Survey (SDSS).
The details of the RCS2 are described in \citet{gilbank2011}.
The small number of filter bands and
the relatively shallow photometry make a mock catalog of RCS2
an excellent test data set for the DEmP method.

All photometric redshift techniques rely on strong features
in the spectroscopic continuum (e.g., the Lyman break, the Balmer break)
to estimate the redshift.
RCS2 has only 4 broad-band photometry,
with the bluest band being $g$, 
which has an effective wavelength of around 4,700\AA, 
with a $\sim1,000$\AA~bandwidth.
Without photometry bluer than $g$,
the $g-r$ color becomes the most important parameter
in estimating the redshifted wavelength of the Balmer break 
(or the 4000\AA~break) for galaxies at $z < 0.4$.
However, an early-type galaxy at $z\sim0.1$
has a $g-r$ color similar to a late-type galaxy at $z\sim0.3$.
Therefore, how to integrate the other minor information
(e.g., magnitudes and colors other than $g-r$) becomes very critical
in breaking the degeneracy of the $g-r$ color for objects at $z < 0.4$,
and hence influence the photometric redshift performance for these objects.
The accuracies of the luminosity and stellar-mass estimations
are also seriously affected by the same issue
since luminosity is highly redshift dependent
and the strength of the 4000\AA~break is a critical parameter
for deriving stellar masses.
Thus, an RCS2 mock catalog is ideal 
for examining the performance of DEmP 
and the comparisons with other methods 
under reasonably challenging circumstances, especially for $z < 0.4$.

The RCS2 mock catalog is generated using the updated version (12/08/2011)
of the COSMOS mock catalog called RealisticSpectroPhotCat 
\footnote {\url{http://lamwws.oamp.fr/cosmowiki/RealisticSpectroPhotCat}.}
\citep{jouvel2009}.
RealisticSpectroPhotCat is generated
by fitting the real COSMOS photometry with a set of SED templates.
The template set includes the Coleman Extended library
(the observed spectra of \citet{cww1980} 
plus extrapolated UV and IR spectra provided by the GISSEL library
(Charlot \& Bruzual 1996)) 
and spectra of star-forming galaxies computed using the GISSEL model.
The Calzetti extinction law \citep{calzetti2000} is used in the fitting.
RealisticSpectroPhotCat provides for over 500,000 objects
both ideal (i.e., modeled) and perturbed (i.e., uncertainty-added) photometry 
in 17 bands from far-ultraviolet to 8.0$\mu$m,
as well as ideal redshifts, stellar masses, 
and star-formation rates.
This mock catalog is representative of a real galaxy survey in many aspects,
such as colors, number counts, and luminosity functions.
Moreover, the detection limits of COSMOS are deeper than those of RCS2,
making the COSMOS mock catalog very suitable for generating a RCS2 mock catalog.

The RCS2 data were obtained
using MegaCam on CFHT with the $g$, $r$, $i'$, and $z'$ filters;
however, the COSMOS mock catalog does not provide photometry for
the CFHT MegaCam filters.
We therefore use the {\it ideal} photometry of 
the Subaru $g$-, $r$-, $i'$-, and $z'$-bands
in the COSMOS mock catalog directly to mimic the CFHT ones.
Directly using the Subaru filter set avoids 
introducing noise from the flux transformations
between the two filter systems.
The transmission curves of the Subaru filters will also be used
in our performance experiment for the template-fitting methods
to make sure there is internal consistency in the filter system.
To match the depths of the RCS2 data,
we added noises and systematics to the mock photometry 
according to the detection limits and the systematic photometric offsets
reported from the real RCS2 data \citep{gilbank2011}.
We then selected objects with $r < 23.5$
to generate the RCS2 mock catalog, which includes $\sim$40,000 objects.

To carry-out out the performance tests,
we need information on redshift, luminosity, and stellar mass
in the RCS2 mock catalog;
but the COSMOS mock catalog provides only redshift and stellar mass 
for each galaxy.
We therefore calculated the luminosities of
the CFHT $u$, $g$, $r$, $i'$, and $z'$ bands (rest) by doing the SED fittings 
using the 17-band photometry with the redshifts 
provided by the COSMOS mock catalog.
We use the EAZY v2.0 code \citep[EAZY, hereafter,][]{brammer2008}
to derive the luminosities of the objects in the RCS2 mock catalog.
These luminosities will be the references 
for examining the qualities of the luminosity estimations
using different methods.

\section{Training Set and Validation Set}\label{trainingset}
The performance of an empirical method would likely be overestimated
if the test set is the same as the training set.
To examine the quality of the empirical method correctly,
an independent validation set should be used.
We therefore separate the RCS2 mock catalog into two sets
by randomly selecting galaxies.
One is the training set 
and the other is the validation set.
Each set contains $\sim$20,000 objects.

As discussed in Section~\ref{balance},
the result of an empirical method can be affected by
the distribution of properties of the galaxies in the training set.
We therefore generated two redshift training sets, 
three luminosity training sets, and three stellar-mass training sets
with different galaxy property distributions, as described below,
to test the different methods for mitigating these effects.

For the empirical redshift fit,
we use two different training sets.
The first training set, $TZ_O$, 
is taken from the RCS2 mock catalog directly, 
which has $\sim$20,000 objects.
The redshift distribution of galaxies of $TZ_O$
is similar to that in the validation set.
For the second training set, $TZ_Z$,
we generate it as a uniformly-weighted training set
using the following procedure.
First, we take the galaxies from $TZ_O$
and separate them into different redshift bins
with a bin size of 0.1 in $z$.
Next, we increase the number of galaxies in each redshift bin
by randomly duplicating galaxies in the same bin
until the number of galaxies in the bin matches 
that in the redshift bin with the largest number of galaxies.
To avoid the singular matrix issue during polynomial fitting,
we apply a small random magnitude offset ($<\pm0.05$ mag)
to each duplicated galaxy.
The magnitude offset is exactly the same for $g$, $r$, $i'$, and $z'$
for the same duplicated galaxy. 
The small offset of the photometry is negligible
for our experiments since the colors of each object are unchanged.
%After this procedure, $TZ_Z$ has $\sim$50,000 objects
%with a flat redshift distribution.

For the empirical luminosity fit, 
we use three different training sets for each rest-wavelength filter band.
The first two training sets, $TLi_O$ and $TLi_Z$
(where $i$ = $u$, $g$, $r$, $i'$, or $z'$ is the filter name),
are the luminosity counterparts of $TZ_O$ and $TZ_Z$.
They are generated following the same procedures
used to create the $TZ_O$ and $TZ_Z$ training sets.
For the third training set, $TLi_L$,
we apply the uniform-weighting to the luminosity distribution
with a 0.5 mag luminosity bin size.
We test five luminosity bands in our experiment,
the total number of luminosity training sets is therefore 15.

For the empirical stellar-mass fit,
we also use three different training sets.
The training sets $TM_O$, $TM_Z$, $TM_M$
are the stellar-mass counterparts of
$TLi_O$, $TLi_Z$, and $TLi_L$, respectively.
The stellar-mass bin size used in the $TM_M$ construction
is 0.5 dex in $log (M_*/M_{\odot})$.

For the experiments using the flattened training sets
(i.e., $TZ_Z$, $TLi_Z$, $TLi_L$, $TM_Z$, and $TM_M$),
because the numbers of objects of these training sets are changed
by the weighting procedure,
we enlarge the size of the local subset of the training set
for the regional polynomial fitting by the same factor
in order to have statistically equivalent local subset sizes.

\section{Results}\label{results}
We examine the performance of the DEmP algorithm in deriving
the two key properties of galaxies, luminosities and stellar mass,
using differently weighted training sets, along with comparing
results from two public available, more traditional template methods.
Since photometric redshift is a key input for any sample selection or
analysis of galaxies, we will first compare the derivations of
photometric redshift from these different methods.
We note that only objects in the validation set are used
for these comparisons.

\subsection{Photometric Redshift}\label{photz}
Photometric redshift is an important input into any analysis of
properties of a photometric sample of galaxies.  Furthermore, they
are input directly into the determination of luminosities and stellar
mass when a template-fitting technique is used. Thus, understanding
the performance of photometric redshift determination will also
inform us in comparing the performance in the derivations of luminosities
and stellar mass.

We examine the performance of DEmP in the determination of photometric 
redshift performance using two weighting schemes for the training set:
no weighting and weighting by redshift distribution.
We also compare the DEmP results with those derived 
using two different template-fitting photometric redshift codes:
New\_Hyperz V11 (NewHyperz, hereafter; Rose et al.
\footnote{\url{http://www.ast.obs-mip.fr/users/roser/hyperz/}}) and EAZY.
NewHyperz is the successor of Hyperz \citep{bmp2000}, 
which is one of the most popular photometric redshift codes
about a decade ago.
It is able to deliver photometric redshifts 
as well as luminosities and stellar masses of galaxies,
but no prior can be applied to improve the performance.
Without applying any prior, 
some objects would have catastrophic photometric redshift errors
due to the confusion between Lyman break and Balmer break
if near-infrared (NIR) data are not available.
Unlike NewHyperz, EAZY can take advantage of an apparent magnitude prior
to reduce the fraction of the catastrophic photometric redshift error.
Moreover, EAZY is able to do SED fits using multiple templates simultaneously.
Therefore, EAZY can be applied to
a galaxy with two or more different stellar populations.
The templates that we used in NewHyperz and EAZY
are the GALAXEV \citep{bc2003} templates 
and the EAZY default template V1.0, respectively.
We use the Calzetti extinction law \citep{calzetti2000} in
both NewHyperz and EAZY.
It is worth noting that
we estimate the zeropoint offset of each filter for NewHyperz and EAZY
by deriving the median difference between 
the photometry in the training set and 
the one estimated using NewHyperz and EAZY with fixed redshifts
provided by the training set.
We then apply the zeropoint offsets to the photometry in the validation sets
for NewHyperz and EAZY.
This may take care of the template mismatch 
between the templates used in NewHyperz/EAZY and those used in the COSMOS mock catalog.
However, we find that the zeropoint offsets are all smaller than 0.02 mag;
the effect due to these small zeropoint offsets is negligible.

\begin{figure*}
\epsscale{1.0}
\plotone{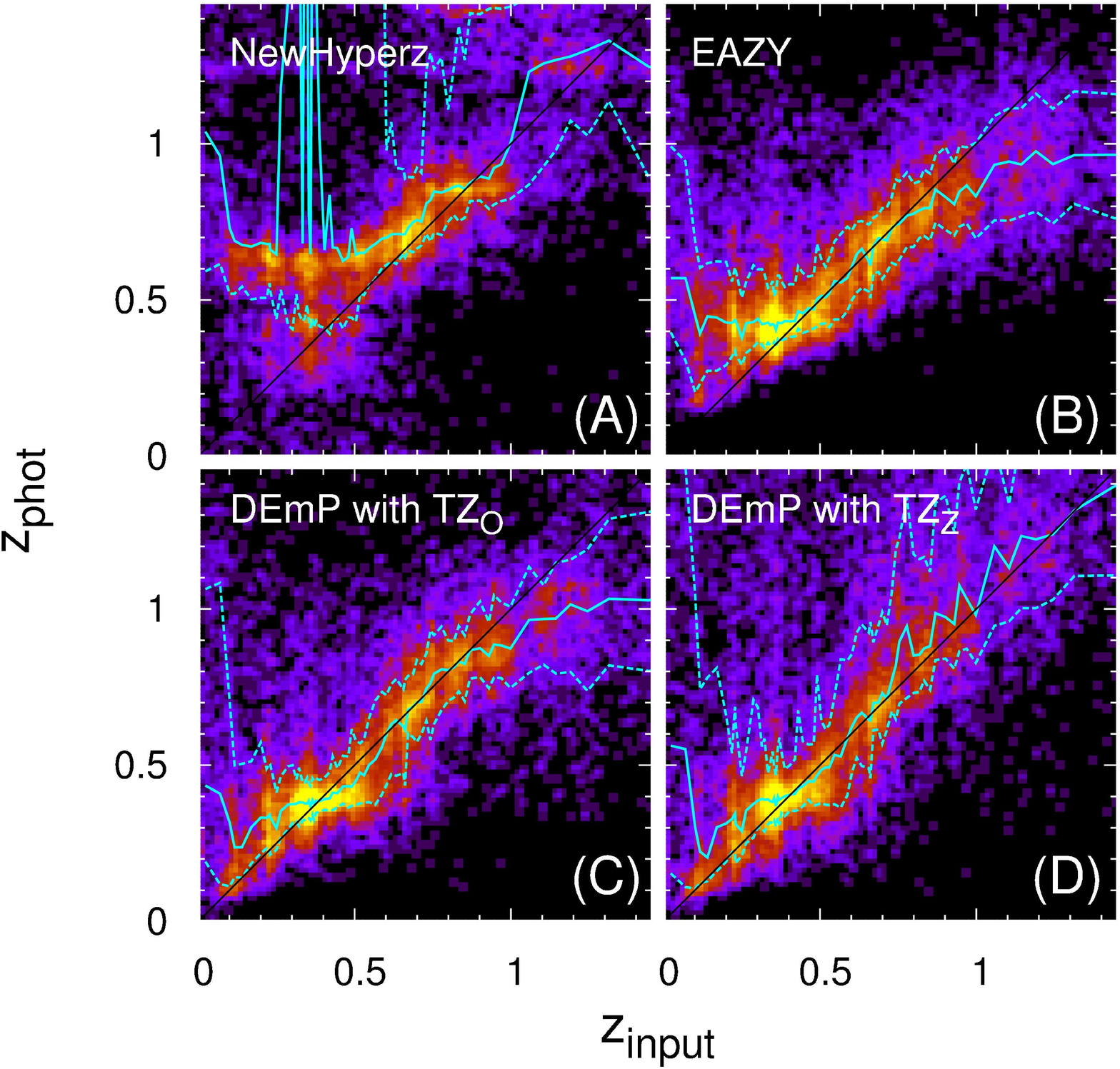}
\caption{The quality of photometric redshift.
Panels A, B, C, and D are for the photometric redshift results
derived using NewHyperz, EAZY, DEmP with $TZ_O$, 
and DEmP with $TZ_Z$, respectively.
The $X$-axis indicates the ideal redshifts in the mock catalog and
the $Y$-axis represents the photometric redshifts 
derived using the different methods.
The solid black line represents the equality of the two redshift variables.
The cyan solid line indicates the running median
and the cyan dashed lines indicate the upper and lower 68th percentiles.
\label{qphotoz} }
\end{figure*}

\begin{deluxetable*}{lrrrrrrrrrrrr}
\tabletypesize{\tiny}
\tablecolumns{13}
\tablewidth{0pt}
\tablecaption{Photometric Redshift Qualities\label{qphotoz_table}}
\tablehead{
\multicolumn{1}{c}{Method} & 
\multicolumn{4}{c}{0.0$<z\leq$0.4} &
\multicolumn{4}{c}{0.4$<z\leq$0.9} &
\multicolumn{4}{c}{0.9$<z\leq$1.5} \\
       & Bias \tablenotemark{a} & 
Scatter \tablenotemark{b} & 
f$_{out}$ \tablenotemark{c} &
f$_{cat}$ \tablenotemark{d} &
Bias & Scatter & f$_{out}$ & f$_{cat}$ &
Bias & Scatter & f$_{out}$ & f$_{cat}$}
\startdata
(A) NewHyperz & 0.884 & 1.177 & 82.6\% & 41.5\% &
0.044 & 0.447 & 30.6\% & 9.4\% &
-0.023 & 0.169 & 40.6\% & 5.1\% \\
(B) EAZY & 0.115 & 0.170 & 41.9\% & 2.1\% &
-0.011 & 0.063 & 6.4\% & 0.4\% &
-0.141 & 0.103 & 36.9\% & 0.6\% \\
(C) DEmP w $TZ_O$ & 0.053 & 0.158 & 24.3\% & 2.7\% &
-0.014 & 0.067 & 9.4\% & 0.7\% &
-0.112 & 0.113 & 30.5\% & 0.8\% \\
(D) DEmP w $TZ_Z$ & 0.041 & 0.182 & 27.2\% & 2.9\% &
-0.005 & 0.088 & 11.5\% & 1.1\% &
-0.011 & 0.130 & 32.2\% & 0.9\% \\
\enddata
\tablenotetext{a}{Median of $z_{phot}-z_{input}$
excluding objects lying outside the running 95th percentile boundary}
\tablenotetext{b}{Standard deviation of
$\frac{z_{input}-z_{phot}}{1+z_{input}} $
excluding objects lying outside the running 95th percentile boundary}
\tablenotetext{c}{Outlier fraction: fraction of objects with 
$\left | \frac{z_{input}-z_{phot}}{1+z_{input}} \right | > 0.15$}
\tablenotetext{d}{Catastrophic error fraction: fraction of objects with
$\left | \frac{z_{input}-z_{phot}}{1+z_{input}} \right | > 0.5$}
\end{deluxetable*}

The comparison results of the photometric redshift performance
are shown in Figure~\ref{qphotoz}.
We also quantify the quality of the photometric redshift 
in three redshift bins for each method
and show the results in Table~\ref{qphotoz_table}.
We note that the bias and scatter of photometric redshift quality
are usually calculated in the literature \citep[e.g.,][]{hildebrandt2010}
excluding outlier objects with the definition
$\left | \frac{z_{input}-z_{phot}}{1+z_{input}} \right | > 0.1$.
However, if the bias of photometric redshift is significant,
calculating bias and scatter with this definition of outlier objects
can produce seriously underestimated results.
Instead, we calculate bias and scatter
excluding objects lying outside the 95th percentile boundary;
the bias and scatter that we derive are therefore better able to
represent the true quality of photometric redshift for each method.
For this paper, we define 0.15 as the outlier boundary.

One may expect the photometric redshifts of objects at $z < 0.4$
to be overestimated because of the absence of $u$-band data
in the RCS2 mock catalog.
In general this is the case.
However, the performance of NewHyperz is significantly worse than 
others over this redshift range.
The running median and upper 68th percentile lines in Panel (A) 
of Figure~\ref{qphotoz} show that
there are many data points outside the plot,
which indicates that the photometric redshifts for objects at $z < 0.5$ 
are seriously overestimated using NewHyperz.
This results in an outlier fraction of 82.6\% for objects at $z < 0.4$,
which is a factor of 2--4 of those of the other methods.
For the intermediate redshift bin, 
NewHyperz still has the worst performance in every aspect.
In the highest redshift bin, 
the performance of NewHyperz is similar to those of the other methods,
except for the catastrophic error fraction.
On the other hand, the performance of EAZY is more reasonable,
with objects with $z < 0.4$ having photometric redshifts around 0.4
and with $z > 0.9$ having photometric redshifts around 0.9.
These results are generally within expectation
because of the lack of $u$-band and NIR data.
The outlier fraction at $z < 0.4$ is only half of that of NewHyperz
because an $r$-band apparent magnitude prior is used in EAZY
to mitigate the catastrophic error problem in our experiment;
it is, however, still a factor of 2 larger than those of the DEmP methods.
For the intermediate redshift bin,
EAZY has the best performance over all the other methods.
But for the highest redshift bin, based on Table~\ref{qphotoz_table},
the bias of EAZY is the most serious,
while the values of the scatter and outlier fraction are
similar to those of the other methods.

While the overall performance of EAZY is reasonably good
given the fact that $u$-band and NIR data are absent,
the overall performance of DEmP is better than that of EAZY.
Using the $TZ_O$ training set, photometric redshifts of objects 
at low redshifts and high redshifts 
are still over and underestimated, respectively,
similar to the EAZY result.
However, the values of bias are milder compared to those of EAZY.
It is also worth noting that 
the bias and the outlier fraction of DEmP with $TZ_O$ at $z < 0.4$ are only 
half of those of EAZY.
For the intermediate redshift bin,
the performance of DEmP with $TZ_O$ is similar to that of EAZY.
We therefore conclude that the performance of DEmP with $TZ_O$
is better than that of EAZY.

The over- and underestimation shown in the DEmP with $TZ_O$ result 
are due to the weighting effect of the non-uniform galaxy distribution
in the redshift space,
and this issue is minimized using DEmP with $TZ_Z$.
Based on Table~\ref{qphotoz_table},
the $\Delta{z}$ bias of DEmP with $TZ_Z$ is the smallest in all redshift bins,
especially at $z>0.9$.
This result shows that the uniformly-weighted training set does reduce the bias
as we expected,
but the price to pay is slightly higher scatter and outlier fraction
as compared to that of DEmP with $TZ_O$.
This is because increasing the weighting of objects
at low-$z$ and high-$z$ makes the fitting function
less optimized for objects at intermediate redshift,
as compared to that derived using $TZ_O$.
For most scientific analyses using photometric redshifts,
the bias and scatter of photometric redshifts 
are more important than the outlier fraction 
and the catastrophic error fraction.
Based on these comparisons, 
we conclude that the photometric redshift performance
of DEmP with $TZ_Z$ is the best in our tested methods.

\subsection{Estimated Luminosities}\label{luminosity}
We compare the performances of the luminosity estimations 
using DEmP and the template-fitting methods.
For the template-fitting method,
we use EAZY to estimate the luminosities by
using the photometric redshifts derived from EAZY and DEmP.
EAZY computes the luminosity of a galaxy
by doing an SED fit using the photometry from the few filters 
with the observed wavelengths close to the redshifted wavelength 
of the chosen filter band.
The derived luminosity is therefore very stable and
not very sensitive to the template used in the SED fitting.
We examine the performances of the different methods in terms of
the luminosity accuracy (expressed in delta absolute magnitude)
as a function of redshift 
and the precision of the reconstructed luminosity function.

\subsubsection{Luminosity versus Redshift}\label{lum}
\begin{figure*}
\epsscale{1.0}
\plotone{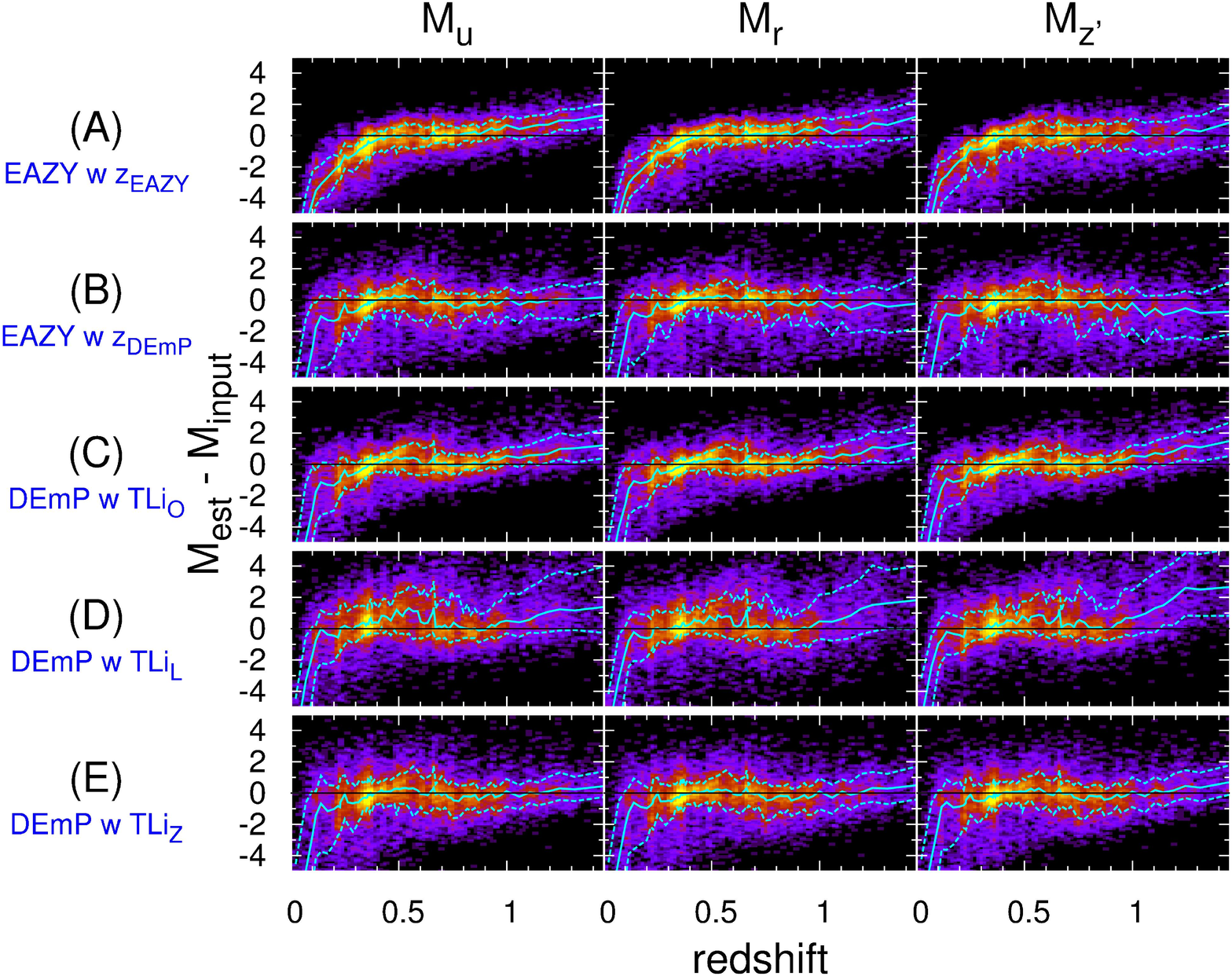}
\caption{Luminosity accuracy as a function of redshift.
Panels from left to right are for 
filter $u$, $r$, and $z'$.
Panels from top (A) to bottom (E) are for
EAZY with EAZY photometric redshifts,
EAZY with DEmP photometric redshifts using the $TZ_z$ training set,
DEmP with $TLi_O$, DEmP with $TLi_L$, and DEmP with $TLi_Z$.
The X-axis in each panel is the ideal redshift in the mock catalog
and the Y-axis is the difference 
between the estimated luminosity and the ideal luminosity 
in the mock catalog in magnitude units.
The cyan solid line indicates the running median
and the cyan dashed lines show the upper and lower 68th percentiles.
Note that the results are very similar for different bands within each method;
for conciseness, the results for the $g$ and $i'$ bands are not shown.
\label{qlum} }
\end{figure*}

We prepare three different luminosity training sets
$TLi_O$, $TLi_L$, and $TLi_Z$ (see Section~\ref{trainingset})
for deriving the luminosities in rest $u$, $g$, $r$, $i'$, and $z'$ bands
using data from observed $g$, $r$, $i'$, and $z'$ bands.
We also use EAZY to derive luminosities in these bands
using the photometric redshifts provided by EAZY and by DEmP.
The photometric redshifts from DEmP used 
in the EAZY template fitting are the ones derived using $TZ_Z$,
since it produces the best photometric redshift performance
according to Section~\ref{photz}.
We find that the performances are similar between
different bands for a given method,
which is demonstrated in Figure~\ref{qlum},
where we show the comparison results for the rest $u$, $r$, and $z'$ bands.
We therefore discuss the general statistics in three redshift bins
for each method only for the $r$ band,
which are tabulated in Table~\ref{qlum_table}.
Because of the lack of $u$-band data, 
the luminosity for objects at $z \leq 0.1$ is seriously overestimated.
The result of the luminosity performance examination 
for objects at $z \leq 0.3$ can be heavily biased by these objects.
We therefore make the lowest redshift bin in Table~\ref{qlum_table}
$0.1 < z \leq 0.3$ and ignore objects at $z \leq 0.1$.

\begin{deluxetable*}{lrrrrrr}
\tabletypesize{\scriptsize}
\tablecolumns{7}
\tablewidth{0pt}
\tablecaption{Luminosity Quality for $M_r$\label{qlum_table}}
\tablehead{
\multicolumn{1}{c}{Method} & 
\multicolumn{2}{c}{0.1$<z\leq$0.3} &
\multicolumn{2}{c}{0.3$<z\leq$0.9} &
\multicolumn{2}{c}{0.9$<z\leq$1.5} \\
       & Bias \tablenotemark{a} & 
Scatter \tablenotemark{b} & 
Bias & Scatter &
Bias & Scatter}
\startdata
(A) EAZY w $z_{EAZY}$ &  -1.07 & 0.72 & -0.22 & 0.53 & 0.44 & 0.77 \\
(B) EAZY w $z_{DEmP}$ &  -0.41 & 0.85 & -0.09 & 0.92 & -0.22 & 1.15 \\
(C) DEmP w $TLr_O$ & -0.51 & 0.77 & 0.11 & 0.54 & 0.42 & 0.73 \\
(D) DEmP w $TLr_L$ & -0.12 & 0.83 & 0.21 & 1.03 & 0.55 & 1.35 \\
(E) DEmP w $TLr_Z$ & -0.35 & 0.84 & 0.04 & 0.74 & 0.12 & 0.79 \\
\enddata
\tablenotetext{a}{Median of $\Delta{M} = M_{estimated} - M_{input}$}
\tablenotetext{b}{Standard deviation of $\Delta{M}$}
\end{deluxetable*}

Because EAZY overestimates redshifts for low-$z$ objects
and underestimates redshifts for high-$z$ objects
as described in \S~\ref{photz},
the luminosity is therefore overestimated for low-$z$ objects
and underestimated for high-$z$ objects
as shown in Table~\ref{qlum_table} and Figure~\ref{qlum} (method A).
Although the luminosity biases for method A at low-$z$ and high-$z$
are relatively large compared to those for the other methods,
its performance in terms of scatter over the entire redshift range
is better than the other methods.
The small luminosity scatter is directly due to 
the small scatter in photometric redshifts derived using EAZY.

Based on the results shown in \S~\ref{photz},
the photometric redshifts derived using DEmP with $TZ_Z$
have the smallest bias among all the methods,
which is also reflected in the bias performance of its luminosity estimation.
The luminosity bias of method B (EAZY with $z_{DEmP}$)
over the entire redshift range is at least a factor of 2 smaller than
that of method A (EAZY with $z_{EAZY}$).
However, the larger photometric redshift scatter of $z_{DEmP}$
is also propagated to its luminosity results,
which makes the luminosity scatter of method B (up to 70\%)
larger than that of method A.

For the template-fitting methods (i.e., methods A and B),
the performance of the luminosity estimation can be easily explained
by the quality of the photometric redshifts used to derive the luminosities.
But for the DEmP methods (i.e., methods C, D, and E),
the same interpretation cannot be applied,
since the luminosities are derived directly from the photometry,
by-passing the photometric redshift estimation.
As discussed in \S\ref{balance}, the result of the empirical method
can be affected by the distribution of galaxy properties in the training set.
For a luminosity training set generated using a flux-limited sample,
objects with intermediate luminosities are more numerous than
those with both fainter and brighter luminosities.
The luminosities of fainter objects are therefore overestimated,
and those with brighter luminosities are underestimated.
Most of these faint objects are at low-$z$
and all the objects at high-$z$ are relatively luminous because of the flux limit.
Hence, the luminosities of low-$z$ objects are systematically overestimated,
while those of high-$z$ objects are underestimated,
as shown in Table~\ref{qlum_table} and Figure~\ref{qlum} (method C).
Although method C still has the overestimation and underestimation issues,
the absolute values of its biases are actually 
between those of the two tested template-fitting methods,
while its scatters are very similar to those of method A,
which has better scatter performance between the two template-fitting methods.
We therefore conclude that the overall performance of method C
is better than those of the template-fitting methods.

\begin{figure*}
\epsscale{1.0}
\plotone{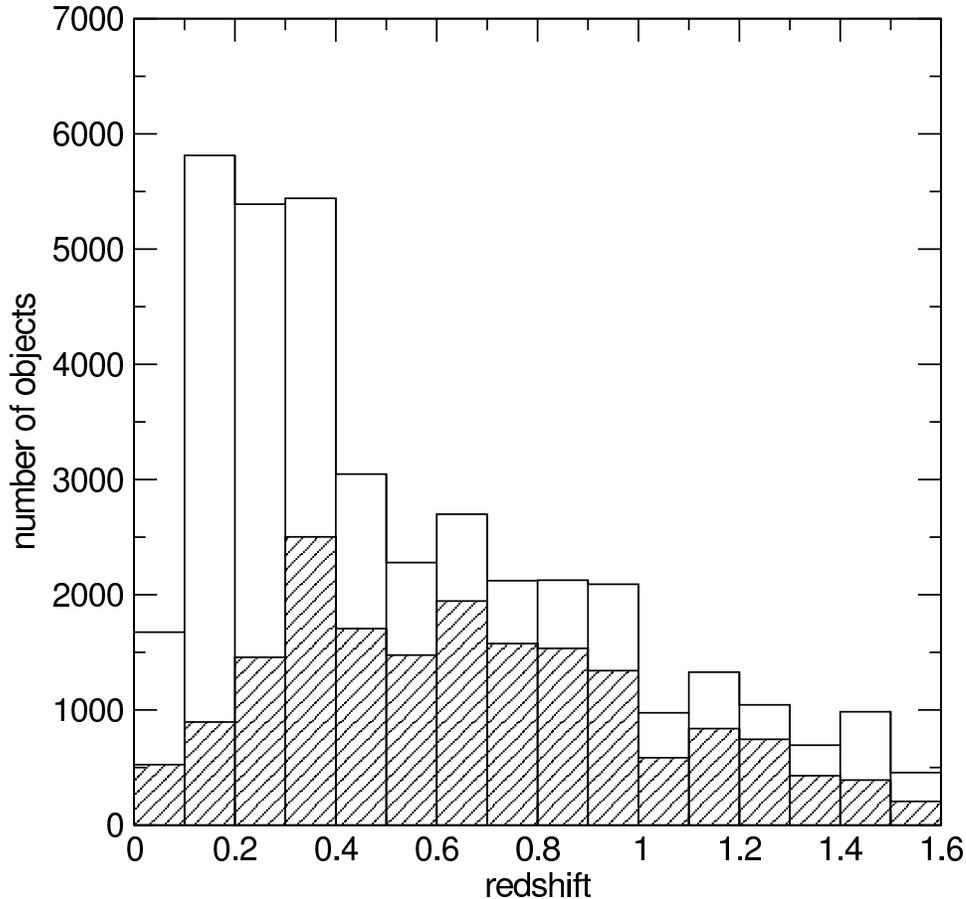}
\caption{Redshift distributions of luminosity training sets:
The hatched histogram shows the redshift distribution 
of the original luminosity training set $TLr_O$,
and the open histogram illustrates the redshift distribution 
for the uniformly weighted luminosity training set $TLr_L$.
\label{lbal_zdist} }
\end{figure*}

Method D uses DEmP with the training set that is
uniformly-weighted in luminosity space.
We expect it to be able to minimize 
the problem of over and underestimating luminosities
at low and high redshifts, respectively,
as it is the case for photometric redshift estimation.
Based on Table~\ref{qlum_table} and Figure~\ref{qlum},
it works as expected for the lowest redshift bin;
the absolute value of the bias at $0.1 < z \leq 0.3$ (0.12) is the smallest
and at least a factor of 3 smaller than those of all the other methods.
However, the bias, as well as the scatter,
for the other two redshift bins are almost the worst among all the methods.
These results can be explained by the weighting factor
applied in obtaining the $TLi_{L}$ training set.
Because the luminosity at the flux limit boundary at higher $z$
changes with redshift much slower than that at lower $z$,
the multiplicative factor applied to each luminosity bin 
for the luminosity uniform-weighting procedure
has a much smaller dependence on redshift at higher $z$.
Hence, the shape of the redshift distribution 
of the uniformly-weighted luminosity training set (i.e., $TLi_L$)
is similar to the original one (i.e., $TLi_O$) for the $z > 0.3$ bins.
To demonstrate this,
we plot the redshift distributions of $TLr_O$ and $TLr_L$ 
in Figure~\ref{lbal_zdist}.
The number of objects in $TLr_L$ with $z \leq 0.4$ is boosted by 
a factor of 1 to 5 compared to that in $TLr_O$;
whereas at $z > 0.4$, the boost is only about 30\%.
The relatively much larger number of objects at in the weighted training set
at low-$z$ produces the best result for the $0.1 < z \leq 0.3$ bin for method D.
Furthermore, since the shape of the redshift distribution of $TLr_L$
is very similar to that of $TLr_O$ at $z > 0.3$,
the bias due to the non-uniform redshift distribution of a training set
for the $z > 0.3$ bins therefore remains.
Moreover, the redshift distribution of $TLr_L$ is significantly shifted
to lower redshift,
so the empirical luminosity fitting can be 
seriously affected by low-$z$ objects (i.e., fainter objects).
This makes the bias and scatter of the estimated luminosity at $z > 0.3$ 
even worse than those of method C (DEmP with $TLi_O$).
In fact, method D produces the largest positive (faint) bias 
in $\Delta{M}$ of all the methods.
Hence, we conclude from the overall performance of method D
that luminosity weighting does not improve the result,
and in fact produces worse outcomes at higher redshift.

Comparing methods D and E, weighting the redshift distribution
produces much better results than applying luminosity weights.
Compared to method C (with training set $TLi_O$),
using the training set $TLi_z$ produces smaller biases in all redshift bins.
There is a slight increase in scatter for all the redshift bins,
but not very significant.
We conclude that method E provides the best overall luminosity results
over the whole redshift range.

\begin{figure*}
\epsscale{1.0}
\plotone{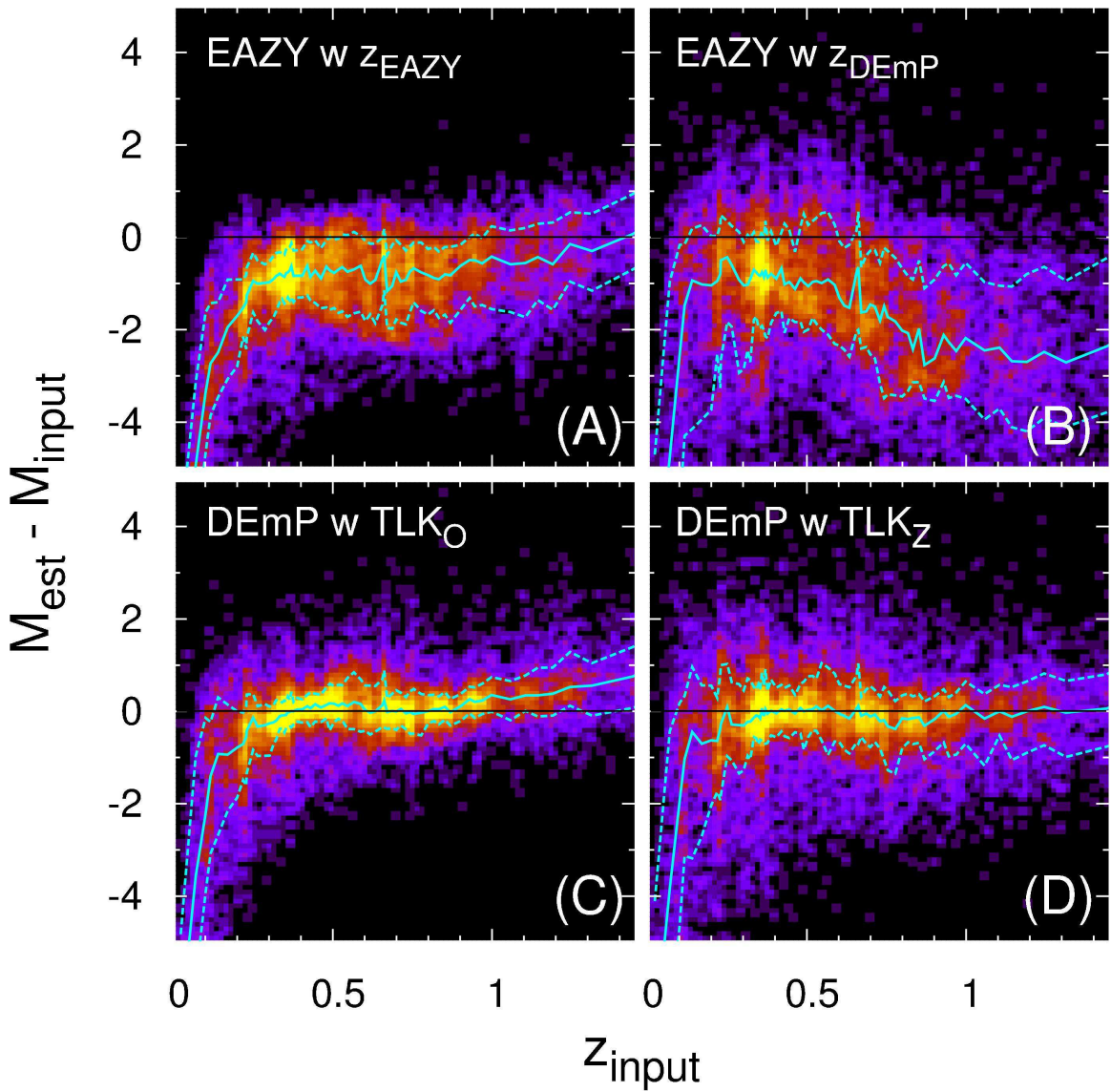}
\caption{$Ks$-band luminosity accuracy as a function of redshift.
Panels A to D are for EAZY with EAZY photometric redshifts,
EAZY with DEmP photometric redshifts, DEmP with the $TLK_O$ training set,
and DEmP with the $TLK_Z$ training set, respectively.
\label{lum_k} }
\end{figure*}

We also find that the performances (especially in terms of scatter) of
the template-fitting methods become worse
when the luminosity band shifts to a longer wavelength,
while those of the empirical-fitting methods do not degrade as much.
For example, the scatter of method B (EAZY with $z_{DEmp}$) 
at $0.9 < z \leq 1.5$ is 0.84 for $u$, but it is 1.29 for $z'$,
while the scatter of method E (DEmP with $TLi_z$)
at $0.9 < z \leq 1.5$ is 0.74 for $u$, and it is 0.82 for $z'$.
To investigate the limitations of each method
to predict luminosities in rest bands not covered by the observed bands,
we repeat the same experiment to estimate the CFHT WIRCam $Ks$-band luminosity
using only photometry from the observed optical bands.
The results are shown in Figure~\ref{lum_k}.
Unlike the results for the optical bands (Figure~\ref{qlum}),
the performances of the estimated $Ks$-band luminosity are significantly different
for the template-fitting and empirical-fitting methods.
For the template-fitting methods,
the bias and scatter of the result derived using EAZY with $z_{DEmP}$
(e.g., bias = -2.7 mag and scatter = 1.4 mag at $z > 0.9$)
are worse than those derived using EAZY with $z_{EAZY}$
(e.g., bias = -0.6 mag and scatter = 0.8 mag at $z > 0.9$),
even though the bias of $z_{DEmP}$ is generally smaller than
that of $z_{EAZY}$ as shown in Figure~\ref{qphotoz}.
This result implies that in extrapolating an SED beyond the observed band 
(e.g., in the optical), 
using photometric redshifts derived with the same SED templates 
may work better than using photometric redshifts derived 
by using other templates or methods, 
even if these other templates or methods produce photometric redshifts 
that are closer to the spectroscopic redshifts.
In addition, the estimated $Ks$-band luminosities
derived using both template-fitting methods are overestimated
by $\sim1$ mag at $0.3 < z < 0.6$. 
On the other hand, for the empirical-fitting methods,
both the bias and scatter performances 
(e.g., bias = 0.06 mag and scatter = 0.81 at $z > 0.3$ for method D)
are very similar to the optical-band luminosities shown in Figure~\ref{qlum}
and are therefore much better than those of the template-fitting methods. 
These results show that the information of $Ks$-band luminosity
is embedded in optical broadband photometry,
and the empirical methods have better ability to squeeze the information out
as compared to the template methods.
The very large improvement of the DEmP methods over
that of the template methods is likely contributed in part
by having $Ks$-band photometry for the training-set galaxies
to provide observational constraints on the $Ks$-band luminosities
extrapolated using optical bands.
The equivalent information (i.e., priors) for the template-fitting method 
is embedded in the templates used. 
However, our test shows that SED template models are clearly not 
as robust as observed priors and is subject to rather large systematics 
when extrapolated to greatly different wavelengths.

\subsubsection{Luminosity Function}\label{lf}
\begin{figure*}
\epsscale{1.0}
\plotone{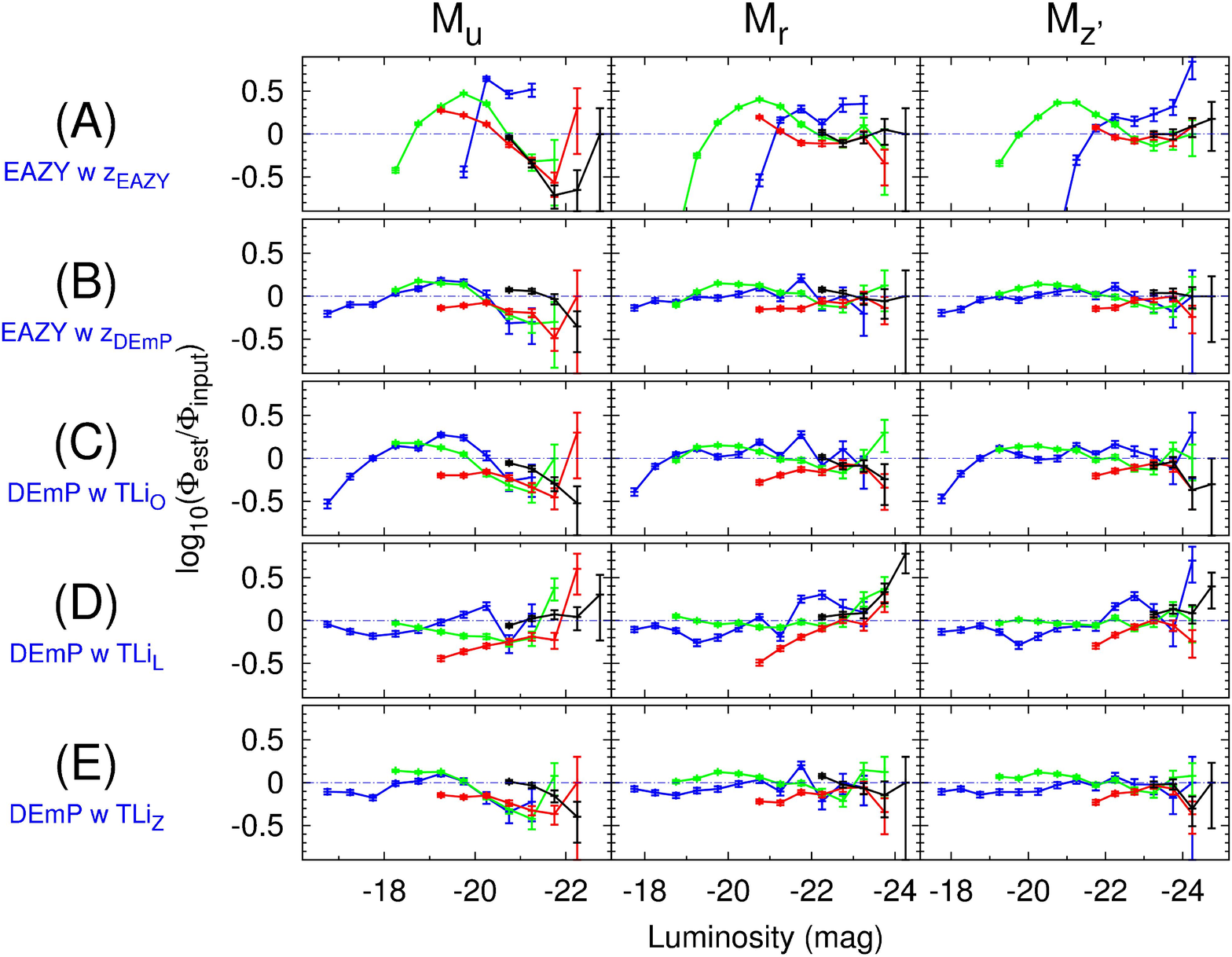}
\caption{Precision of the luminosity function construction.
The method used for each panel is the same as that in Figure~\ref{qlum}.
The X-axis is the luminosity for each band
and the Y-axis is the logarithmic volume-density ratio between
the constructed luminosity function and the ideal luminosity function
per luminosity interval.
The blue, green, red, and black lines
are for 0.0--0.3, 0.3--0.5, 0.5--0.7, and 0.7--0.9 redshift bin, respectively.
\label{qlf} }
\end{figure*}

We examine how well luminosity functions at different redshift ranges
can be reconstructed using different methods.
Since the COSMOS mock catalog is built directly 
from the observed COSMOS catalog,
the COSMOS mock catalog preserves the information of the luminosity functions
in the Universe to the detection limit of the COSMOS data.
To mimic real data, 
we use the derived photometric redshifts 
to populate the redshift bins in which luminosity functions are measured.
The photometric redshifts for method A are derived using EAZY
and those for methods B, C, D, and E
are derived using DEmP with $TZ_Z$.
The completeness limit of the luminosity function for each redshift bin
and for each band is estimated by comparing the luminosity function
constructed using the ideal redshift and the ideal luminosity
of the RCS2 sample (i.e., training set plus validation set)
and that constructed using the whole COSMOS sample.
The results are shown in Figure~\ref{qlf}.
There are four redshift bins:
0.0--0.3, 0.3--0.5, 0.5--0.7, and 0.7--0.9.
Only data points within the completeness limit are shown.
The luminosity function at $z > 0.9$ is mostly incomplete
and therefore not shown in this figure.

We find that the quality of a derived luminosity function is tightly related to 
the biases of both estimated luminosity and photometric redshift.
The quality of the luminosity function is also wavelength dependent:
For all methods, the quality for redder filters is better than
that for bluer filters.
Method A (EAZY with $z_{EAZY}$) uses the luminosity and photometric redshift
that have the best performance in terms of scatter
among all the methods in general,
but the quality of its luminosity functions is the worst.
For the $u$-band luminosity functions, 
the bright-ends are underestimated by 0.5 dex,
the intermediate parts are overestimated by 0.5 dex,
and the faint-ends are seriously underestimated again.
For the luminosity functions of redder bands,
the quality becomes better, especially for the bright-ends.
The faint-ends, however, are still seriously underestimated
as compared to those derived using the other methods.
This poor performance is primary due to
the relatively large biases of $z_{EAZY}$ and luminosity estimates at $z < 0.3$.

Compared to method A, 
the performance of method B (EAZY with $z_{DEmP}$) is much better,
even though both of them are template-fitting methods.
While the scatters of the photometric redshift ($z_{DEmP}$)
and the luminosity derived using method B 
are larger than those from method A,
they both have significantly smaller biases compared to those of method A
(see Tables~\ref{qphotoz_table} and \ref{qlum_table}).
The luminosity function comparison results suggest that
the bias performance of photometric redshift and luminosity 
is the key to the quality of the recovered luminosity functions;
whereas the scatters in these quantities have relatively minor consequences.

For the luminosity functions derived using the DEmP method
(C, D, and E, using $TLi_O$, $TLi_L$, and $TLi_Z$, respectively),
we use $z_{DEmP}$ for the purpose of binning the redshift,
as is the case for method B.
The differences in the qualities of the luminosity functions derived
using these methods are therefore completely due to
the differences in the estimated luminosities.
According to Table~\ref{qlum_table}, 
the ranking of the performance of the luminosity bias from better to poor is:
(E) $>$ (C) $>$ (D),
which agrees with the performance comparison of the luminosity functions
shown in Figure~\ref{qlf}.
This supports the argument that luminosity bias is more important than
luminosity scatter for luminosity function recovery.
For the lowest redshift bin,
the biases of the faint-ends of the luminosity functions for method C
(DEmP with $TLi_O$) are larger than those for methods D and E.
This is the result of the number of faint galaxies at low redshifts in $TLi_O$
being much less than that of brighter galaxies;
the luminosities of these faint galaxies are therefore overestimated.
Either weighting the training set in luminosity space (i.e., $TLi_L$)
or in redshift space (i.e., $TLi_Z$) can 
increase the number of these faint galaxies at low redshifts
in the training set;
the luminosity biases of these galaxies
for methods D and E are therefore minimized.
Since the luminosity function results for methods B and E are very similar,
we conclude that both of these two methods have the best performance
in this experiment.

\subsection{Estimated Stellar Masses}\label{mass}
We compare the estimated stellar masses using DEmP
with those using the conventional SED fitting methods in this section.
For the template-fitting methods, 
we use NewHyperz with the GALAXEV \citep{bc2003} templates 
to derive stellar masses,
applying the photometric redshifts derived using 
NewHyperz, EAZY, and DEmP with $TZ_Z$.
As has been done for the luminosity experiment,
we examine the performances of the stellar-mass estimations
by checking the accuracy of the derived stellar mass
as a function of redshift
and the precision of the constructed stellar-mass function.

\subsubsection{Stellar Mass versus Redshift}\label{mass_z}
\begin{figure*}
\epsscale{1.0}
\plotone{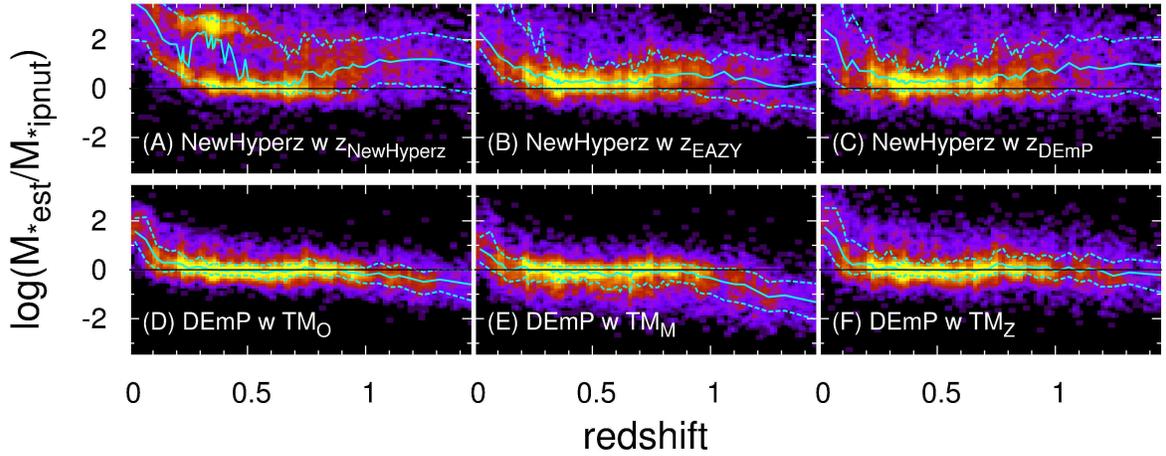}
\caption{Estimated stellar-mass accuracy as a function of redshift.
Panels A, B, C, D, E, and F are for
NewHyperz with the NewHyperz photometric redshifts,
NewHyperz with the EAZY photometric redshifts,
NewHyperz with the DEmP photometric redshifts, 
DEmP with $TM_O$, DEmP with $TM_M$, and DEmP with $TM_Z$, respectively.
The X-axis is the ideal redshift from the mock catalog and
the Y-axis is the logarithmic of the ratio of the estimated mass to the input (ideal) mass.
The cyan solid line indicates the running median
and the cyan dashed lines indicate the upper and lower 68th percentiles.
\label{qmass} }
\end{figure*}

We show the accuracy of the estimated stellar mass
as a function of redshift for each method in Figure~\ref{qmass}
and the statistics of these results in Table~\ref{qmass_table}.
We exclude objects at $z \leq 0.1$ in Table~\ref{qmass_table}
based on the same reason described in \S~\ref{lum} 
for Table~\ref{qlum_table}.

For the template-fitting methods (methods A, B, and C),
the qualities of photometric redshifts used to derive stellar masses
are directly reflected in the estimated stellar masses.
$z_{NewHyperz}$ has serious catastrophic errors at low-$z$
and overestimation at high-$z$,
which are also apparent in the estimated stellar-mass result
(panel A of Figure~\ref{qmass}).
$z_{EAZY}$ has smaller scatter compared to $z_{DEmP}$;
the scatter of the estimated stellar masses in panel B
(NewHyperz with $z_{EAZY}$) is smaller than that in panel C
(NewHyperz with $z_{DEmP}$).
However, one behavior shown in the photometric redshift results 
(see Figure~\ref{qphotoz} and Table~\ref{qphotoz_table})
is not propagated to the estimation of stellar mass.
$z_{DEmP}$ has smaller biases than $z_{EAZY}$ at $z \leq 0.4$,
but the biases of the estimated stellar masses for methods B and C,
using $z_{EAZY}$ and $z_{DEmP}$, respectively, are similar.
In addition, the estimated stellar masses derived using
the template-fitting methods are all overestimated 
by roughly 0.4 dex at $0.3 < z \leq 0.9$.
Figure~\ref{lum_k} shows that template-fitting methods overestimate
the $Ks$-band luminosity by one magnitude.
If we assume that the $Ks$-band luminosity 
is directly correlated with stellar mass,
a one magnitude difference in $Ks$-band luminosity translates to
a 0.4 dex difference in stellar mass, which is consistent with
the stellar-mass biases derived using the template-fitting methods.

For the empirical-fitting methods (methods D, E, and F),
the quality of the estimated stellar mass cannot be simply explained
by the quality of the photometric redshift,
since the stellar mass is derived directly from the photometry,
by-passing photometric redshift determination. 
For a stellar-mass training set constructed from a flux-limited database,
the mean stellar mass for galaxies at low-$z$ in the training set
is lower than that for galaxies at intermediate redshift,
and the number of objects at low-$z$ is smaller than 
that at intermediate redshift.
Similarly, the mean stellar mass for galaxies at high-$z$
is higher than that for galaxies at intermediate redshift,
and the number of objects at high-$z$ is lower than
that at intermediate redshift as well.
Hence, the stellar masses of objects at $z \leq 0.3$
are biased by the objects at intermediate redshift and overestimated,
and those at $z > 0.9$ are underestimated because of the same reason.
This effect can be seen in Figure~\ref{qmass} for method D (DEmP with $TM_O$).

Method E uses the training set that is uniformly-weighted
in stellar-mass space ($TM_M$).
Although the stellar-mass distribution in $TM_M$ is flat,
the redshift distribution is seriously distorted,
similar to that for the $TLi_L$ training set (see Figure~\ref{lbal_zdist}).
The uniform stellar-mass weighting process 
boosts the number of objects at $z \leq 0.3$ by a factor of $\sim4.2$ 
while the number of objects at the other redshift range
just increases by a factor of $\sim1.6$.
This reduces the bias at $z \leq 0.3$,
but increases the scatter at $z > 0.3$ and the bias at $z > 0.9$,
as compared to those of method D (DEmP with $TM_O$).

Method F uses the training set uniformly-weighted
in redshift space ($TM_Z$).
Because the redshift distribution in $TM_Z$ is flat,
the stellar-mass biases at $z \leq 0.3$ and $z > 0.9$ are reduced
as compared to those of method D (DEmP with $TM_O$).
Its stellar-mass bias at $z \leq 0.3$ is not as small as that of method E (DEmP with $TM_M$),
but the performance at $z > 0.3$ is much better than that of method E
in terms of both bias and scatter.
Therefore, method F delivers the best result among the empirical methods.

Generally speaking, the results derived using the empirical methods
are all better than those derived using the template methods
in the experiment of stellar-mass estimation.
In conclusion, method F (DEmP with $TM_Z$)
has the best balance between bias and scatter,
outperforming all other tested methods.

\begin{deluxetable*}{lrrrrrr}
\tabletypesize{\scriptsize}
\tablecolumns{7}
\tablewidth{0pt}
\tablecaption{Stellar-Mass Qualities\label{qmass_table}}
\tablehead{
\multicolumn{1}{c}{Method} & 
\multicolumn{2}{c}{0.1$<z\leq$0.3} &
\multicolumn{2}{c}{0.3$<z\leq$0.9} &
\multicolumn{2}{c}{0.9$<z\leq$1.5} \\
       & Bias \tablenotemark{a} & 
Scatter \tablenotemark{b} & 
Bias & Scatter &
Bias & Scatter}
\startdata
(A) NewHyperz w $z_{NewHyperz}$ & 1.88 & 1.42 & 0.52 & 1.57 & 0.96 & 1.02 \\
(B) NewHyperz w $z_{EAZY}$ & 0.68 & 1.03 & 0.39 & 0.66 & 0.47 & 0.89 \\
(C) NewHyperz w $z_{DEmP}$ & 0.65 & 0.97 & 0.36 & 0.81 & 0.93 & 1.00 \\
(D) DEmP w $TM_O$ & 0.18 & 0.41 & -0.02 & 0.22 & -0.25 & 0.45 \\
(E) DEmP w $TM_M$ & 0.05 & 0.49 & -0.08 & 0.46 & -0.39 & 0.72 \\
(F) DEmP w $TM_Z$ & 0.19 & 0.61 & 0.01 & 0.34 & -0.07 & 0.53 \\
\enddata
\tablenotetext{a}{Median of $\Delta{log(M_{star})}
= log(M_{star,estimated}) - log(M_{star,input})$}
\tablenotetext{b}{Standard deviation of $\Delta{log(M_{star})}$}
\end{deluxetable*}

%\begin{figure*}
%\epsscale{1.0}
%\plotone{mbal_zdistribution.eps}
%\caption{Redshift distributions of stellar mass training sets.
%The hatched histogram indicates the redshift distribution of $TM_O$
%and the open histogram indicates the redshift distribution of $TM_M$.
%\label{mbal_zdist} }
%\end{figure*}

\subsubsection{Stellar-Mass Function}\label{mf}

\begin{figure*}
\epsscale{1.0}
\plotone{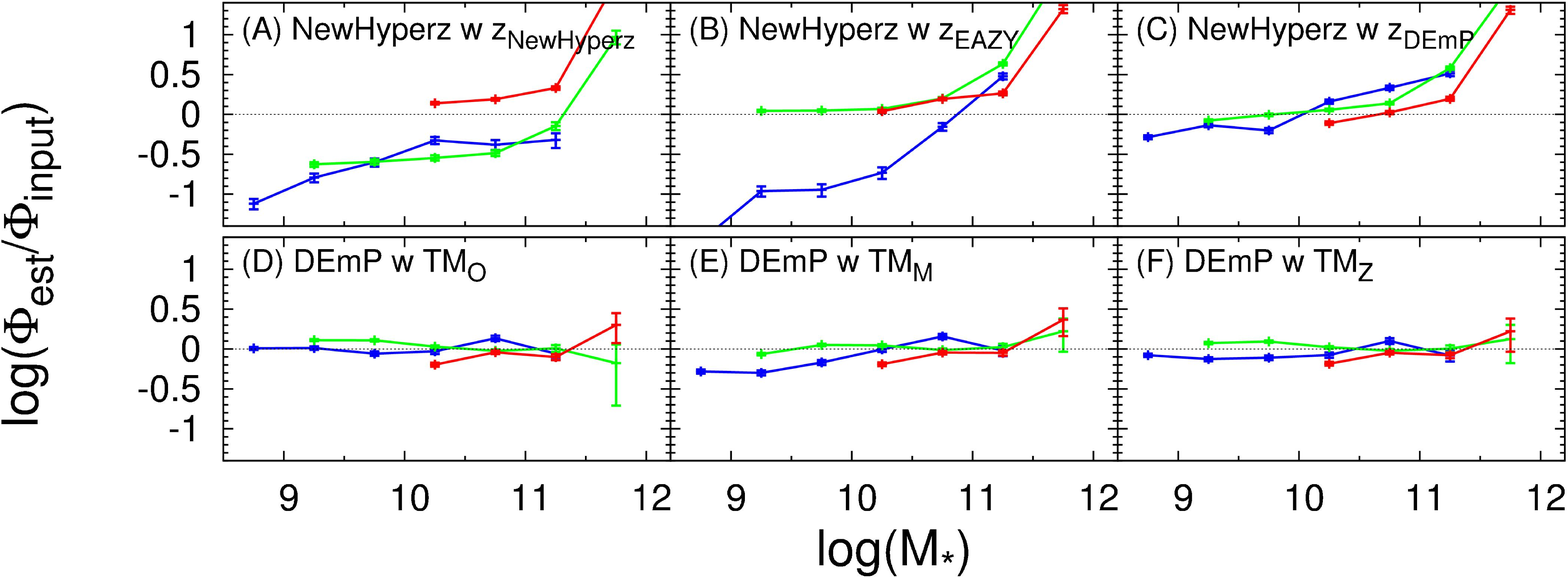}
\caption{Precision of the reconstructed stellar-mass function.
The method used for each panel is the same as that in Figure~\ref{qmass}.
The X-axis is the stellar mass in logarithmic scale
and the Y-axis is the logarithmic volume-density ratio between
the stellar-mass function constructed from the estimated stellar mass
and the ideal stellar-mass function per stellar-mass interval.
The blue, green, and red lines
are for 0.0--0.3, 0.3--0.5, and 0.5--0.7 redshift bins, respectively.
\label{qmf} }
\end{figure*}

We examine how well the stellar-mass function 
at different redshift ranges can be constructed using the different methods.
As has been done in the luminosity function experiment,
we use the derived photometric redshifts to populate the redshift bins 
for the stellar-mass functions to mimic real data.
The photometric redshifts used for the redshift bins in method A 
are derived using NewHyperz,
those in method B are derived using EAZY,
and those in methods C, D, E, and F are derived using DEmP with $TZ_Z$.
The completeness limit of the stellar-mass function for each redshift bin
is estimated by comparing the stellar-mass function
constructed using the ideal redshift and the ideal stellar mass
of the RCS2 sample and that constructed using the whole COSMOS sample.
The results are shown in Figure~\ref{qmf}.
There are three redshift bins:
0.0--0.3, 0.3--0.5, and 0.5--0.7.
Only data points within the completeness limit are shown.
The stellar-mass function at $z > 0.7$ is mostly incomplete
and therefore not shown in this figure.

Figure~\ref{qmf} shows that
all the results derived using the template-fitting methods
have significant biases.
These biases are in general due to two reasons:
the biases of photometric redshifts 
which move objects to wrong redshift bins 
and also lead to biased stellar-mass estimates;
and the significant stellar-mass biases introduced by the template-fitting methods,
as described in \S\ref{mass_z}.

For method A (NewHyperz with $z_{NewHyperz}$),
the stellar-mass functions for $0.0 < z \leq 0.3$ and $0.3 < z \leq 0.5$
are significantly underestimated ($\sim$ 0.5 dex)
for objects with stellar mass less than $10^{11}$ M$_\odot$.
This is the result of the significantly overestimated photometric redshifts
for objects at $z < 0.4$ derived using NewHyperz.
Since objects at these redshift bins are offset or scattered
to the other higher redshift bins,
the numbers of objects in these redshift bins decrease significantly,
resulting in the underestimate of the stellar-mass functions.
Meanwhile, the stellar masses of these scattered objects are overestimated
because of their overestimated photometric redshifts;
the numbers of massive objects ($> 10^{11}$ M$_\odot$) 
in the higher redshift bins
(i.e., the $0.3 < z \leq 0.5$ and $0.5 < z \leq 0.7$ bins)
therefore increase significantly,
resulting in the overestimated stellar-mass functions
at the high-mass end.

Method B uses $z_{EAZY}$ to reconstruct the stellar-mass functions.
For $z \leq 0.3$, most galaxies have their redshifts scattered to greater than 0.3,
causing the stellar-mass function for the lowest redshift bin 
to be seriously underestimated.
These objects with overestimated photometric redshifts 
also have their stellar masses overestimated as well. 
In addition, the biases of the estimated stellar masses
(see method B in Table~\ref{qmass_table})
also shift the stellar-mass function to the massive end.
These effects lead to the overestimated high-mass ends of 
the stellar-mass functions for the 0.3--0.5 and 0.5--0.7 redshift bins. 

Method C uses $z_{DEmP}$ to generate the mass functions.
The redshift bias of $z_{DEmP}$ is smaller than $z_{NewHyperz}$ and $z_{EAZY}$,
especially for $z < 0.4$;
the stellar-mass function for the lowest redshift bin 
is therefore not as seriously underestimated
as those derived using methods A and B.
However, the bias in the estimated stellar masses
(see method C in Table~\ref{qmass_table})
makes the stellar-mass function offset to the massive end;
the high-mass ends of the stellar-mass functions for all redshift bins
are therefore overestimated.
Based on Figure~\ref{qmf},
method C outperforms all the other template-fitting methods.

For the empirical-fitting methods (methods D, E, and F),
the photometric redshift used to reconstruct the stellar-mass functions
is $z_{DEmP}$, which has the least photometric redshift bias.
Furthermore, the stellar masses estimated using the empirical-fitting methods
have relatively low systematic bias.
Therefore, not surprisingly, the qualities of the estimated stellar-mass functions
derived using the empirical-fitting methods are better than
those derived using the template-fitting methods.
The stellar mass used in method E is derived using $TM_M$
(see \S~\ref{mass_z}), and has the largest scatter and bias
among the empirical-fitting methods.
The performance of the recovered stellar-mass functions is correspondingly
the worst amongst the empirical methods, but not by much;
the variations of the mass functions in all the redshift bins
are within $\pm0.4$ dex.
On the other hand, method F uses the stellar mass derived using $TM_Z$,
which has the best overall performance in mass estimation.
As shown in Figure~\ref{qmf}, panel F,
the variations of the stellar-mass functions in all the redshift bins
are within $\pm0.2$ dex.
The performances of methods D and F are similar.
Method D, which uses the unweighted training set $TM_O$, 
produces stellar masses with biases similar or slightly larger 
than those of Method F in all the redshift bins, 
but with somewhat smaller scatters (see Table~\ref{qmass_table}). 
It generates stellar-mass functions 
with performance very similar to those of Method F.  
We therefore conclude that Methods D and F provide the best results 
in reconstructing the stellar-mass function amongst all the methods tested.

\section{Discussion}\label{discussion}
Our analysis using a mock catalog of four-band photometry
has shown that DEmP with the uniform redshift-weighted training sets 
outperforms all the other methods and training sets in our experiments.
The results show that deriving luminosities and stellar masses
from photometry directly (i.e., bypassing the photometric redshift estimation)
delivers significantly improved results.
This is the result of the DEmP method avoiding introducing noises
caused by photometric redshift errors and template uncertainties.
These mock catalog experiments also show that
resampling the training set galaxy sample 
to have a uniformly weighted redshift distribution 
produces the best results in not just estimating photometric redshift, 
but also luminosity and stellar mass.

Besides the performance comparisons, 
the results show some other interesting phenomena that we did not expect.
One may naively expect that 
one can only derive absolute luminosities for bands
that are bracketed by the observed bands
since one can interpolate across the SED between the observed bands.
If the redshifted wavelength of the selected band
is longer than the wavelength of the reddest filter
or shorter than that of the bluest filter,
the estimate of the luminosity is expected to be less reliable.
Therefore, with the RCS2 filter configuration,
we expect the luminosity of a given band to be reliable
only within a certain redshift range:
$0.3 < z < 1.6$ for the $u$-band luminosity;
$0.0 < z < 1.0$ for the $g$-band luminosity;
$0.0 < z < 0.5$ for the $r$-band luminosity;
and $0.0 < z < 0.2$ for the $i'$-band luminosity.
The $z'$-band luminosity for any redshift can only be derived by extrapolation.
However, Figures~\ref{qlum} and \ref{qlf}
suggest that the performances of the empirical luminosity estimations
are very similar for the different bands within each of the DEmP methods.
Surprisingly, Figure~\ref{lum_k} shows that 
the performances of the DEmP $Ks$-band luminosity estimations
are also very similar to those of the optical luminosity estimations.
These results imply that the observed NIR photometry at $< 2.5\mu$m for galaxies 
at $z < 1.2$ can be well constrained 
by their observed multi-wavelength optical photometry only,
assuming one has near-infrared photometry for the training set.
Thus, in general, obtaining a relatively small set of training set data
in filter bands other than those in the main data set
will allow one to extrapolate photometric information for these bands
using an empirical-fitting method.

The results of our stellar-mass experiment also
show some unexpected interesting results.
Conventional wisdom suggests that NIR data are required
for deriving stellar masses of galaxies,
since most of the NIR photons are emitted from low-mass stars
which dominate a great part of the total stellar mass of a galaxy.
Furthermore, NIR luminosities are less affected by 
star-formation rate (SFR) and dust extinction.
Thus, it is generally accepted that the NIR luminosity 
is a better stellar-mass proxy than the optical luminosity.
However, based on Figures~\ref{qmass} and \ref{qmf},
we find that DEmP is able to estimate stellar mass relatively accurately,
using only optical filter bands up to 9200\AA.
It is not surprising that DEmP is able to estimate
the stellar mass reasonably well given that
it can derive near-infrared luminosities 
with a similar accuracy from optical data.
The small set of stellar masses derived for the training set
provides the additional priors needed to anchor the stellar mass estimates.
Some previous studies also found that
the stellar mass of galaxies can be reasonably well derived
using multi-wavelength optical broad-band photometry
\citep[e.g.,][]{bd2001,bell2003,zcr2009,taylor2010,taylor2011},
which is consistent with the results of our experiments.
However, it is worth noting that all these papers
use spectroscopic redshifts in their measurements.
For a large sky survey, 
it is very time-consuming to obtain spectroscopic redshifts for all the sources.
DEmP appears to be an excellent and robust method for estimating stellar 
masses for a large optical broadband survey.

A conventional way to estimate the stellar mass 
using a template-fitting method for a galaxy with photometric redshift 
is to derive its redshift using a template 
that can deliver the most accurate photometric redshift, and then, 
using this photometric redshift, 
to derive the stellar mass using a template SED 
that provides information on stellar mass \citep[e.g.,][]{muzzin2013}.
However, templates providing information on stellar mass
(e.g., GALAXEV) are usually not the best template 
for measuring photometric redshifts.
If different templates are used for estimating redshifts and stellar masses,
the SED for deriving the stellar mass may not fit the photometry properly 
for the given fixed photometric redshift.
This can cause larger scatter and bias in the estimated stellar mass.
For example, both the photometric redshifts and the stellar masses
in panel A of Figure~\ref{qmass} are derived using NewHyperz
with the GALAXEV template.
This method produces a large number of catastrophic photometric redshift errors 
(41\%, see Table~\ref{qphotoz_table}) at $z < 0.4$ 
due to the lack of absolute magnitude priors. 
As a result, a large number of galaxies at $z < 0.4$ have very large errors 
in their stellar-mass estimates, 
as shown in the upper right part of Panel A in Figure~\ref{qmass}
However, inspecting the dat points in Panels A, B, and C 
in Figure~\ref{qmass} for galaxies with $0.3 < z < 0.7$, 
it can be seen that galaxies which do not suffer from catastrophic errors 
have smaller bias and scatter in their stellar mass estimations 
than those for Methods B and C which use different templates or methods 
for estimating redshifts and stellar masses. 
Moreover, the best photometric redshift for a certain object is usually
calculated from its probability distribution function of photometric redshift
instead of the one with the minimum $\chi^2$.
Therefore, even when the same template is used for
estimating redshifts and stellar masses,
the SED fitting for the stellar-mass estimation may not be optimal
if the best photometric redshift is used.
Deriving stellar masses using DEmP can avoid these issues.

One of the main reasons that an empirical method such as DEmP 
can provide more robust results with smaller bias is that 
the training set basically provides a set of priors 
for the quantity that one wants to measure. 
Compared to template-fitting methods, 
this set of priors is equivalent to the templates. 
However, having a set of measured priors appears to produce 
better results than using a set of model templates.

The quality of the results delivered by DEmP
relies on the quality of the training set.
A good training set needs to cover a larger volume
in the multi-dimensional magnitude-color space than the target set
in order to provide a good solution for every object in the target set.
In other words, a good training set has to be more complete 
than the target set in terms of all the related aspects,
such as depth, galaxy type, and redshift, etc.
However, it is very difficult to generate a good training set.
For example, the typical success rate in determining redshifts
from spectroscopic data is about 70\% for a flux-limited sample.
This means about 30\% of galaxies are not included in the training set, 
primarily due to the lack of strong features in their spectra. 
These missing galaxies usually belong to certain types of galaxies.
Thus, the DEmP result for these unsampled populations can be unreliable.
One possible solution is to use template fitting with photometric data 
consisting of a great number of bands (e.g., more than 20 bands) 
to assess these unsampled populations 
and to account for the incompleteness of the spectroscopic training set. 
However, the uniformity of the training set can be affected 
if some of the members of the training set are obtained using a different method.

There may be 10- or even 15-band large sky surveys in the future.
The comparison of the performances of template-fitting methods
and the DEmP methods for data with more than 10 bands
can be very different from those shown in this paper.
However, we focus only on how to derive the best results
from a dataset with only a few bands in this paper,
and find that DEmP is the more powerful method under this condition.
Investigating how well the template-fitting methods and DEmP perform
using a dataset with more than 10 bands is out of the scope of this paper.

\section{Summary}\label{summary}
In this paper, we introduce a simple but robust method, DEmP,
which is able to deliver redshifts, luminosities, and stellar masses
from multi-wavelength photometry empirically.
It has two important features 
to minimize the two major issues in conventional empirical-fitting methods.
First, DEmP uses a local subset of the training set 
with the 50 nearest neighbors
in the multi-dimensional color-magnitude space for each galaxy
to derive the photometric redshift for that galaxy.
This feature addresses the issue of the suitability 
of the form for the empirical function 
by using a simple function applied locally.
Second, DEmP applies weighting to the objects in the training set
to minimize the bias effect due to the non-uniformity of training set
galaxy properties such as redshift, luminosity, and stellar mass.

Based on the performance tests using an RCS2 mock catalog,
we find that in general DEmP with a training set uniformly weighted
in redshift space provides the best results in deriving 
photometric redshifts, when compared to DEmP with no training set
weighting, or the template fitting methods.
Somewhat surprisingly, this conclusion regarding 
a redshift-weighted training set also applies to DEmP used 
for the direct derivations of galaxy luminosities and stellar mass,
bypassing the use of photometric redshift;
it outperforms other DEmP methods with
unweighted training sets, or training sets weighted to have uniform
luminosity or stellar-mass distributions.
It also similarly outperforms the template fitting methods.

The DEmP method can also accurately estimate luminosities in rest 
filter bands well outside the range of the observed bands in the data.
This is demonstrated by using the 4 observed optical bands in the mock catalog 
to derived Ks band luminosities, 
producing results that are more than a factor of two more accurate than 
those from template fitting methods.
This improved result is likely in part due to having Ks-band data in the
training set which serve the purpose of being a prior in the estimate.
A similar conclusion can be drawn for estimating stellar masses, since
they are fairly closely related to NIR photometry.

In addition to redshift, luminosity and stellar mass,
DEmP can also be applied to derive other intrinsic properties of galaxies.
For example, by assuming SFR and age are functions of photometry,
one is able to estimate the SFRs and ages of galaxies empirically.
The SFR and age training sets can be constructed using
the SFRs and ages estimated from suitable spectroscopic datasets.
In fact, studies in all aspects of galaxy evolution can benefit from DEmP,
as long as the output value can be assumed to be a function of the inputs,
delivering more robust results than those using conventional template-fitting methods.

\end{document}